\documentclass[showpacs,preprintnumbers,prl,twocolumn]{revtex4}
\usepackage{amsmath}
\usepackage{graphicx}
\usepackage{dcolumn}
\usepackage{bm}

\input{tcilatex}

\begin{document}

\title{Magnetic dipole--vortex interaction in a bilayer
superconductor/soft-magnet heterostructure}
\author{S.~V.~Yampolskii}
\altaffiliation[On leave from ]{Donetsk Institute for Physics and Technology, National Academy of Sciences
of Ukraine, 83114 Donetsk, Ukraine. Electronic address:
yampolsk@tgm.tu-darmstadt.de.}
\author{G.~I.~Yampolskaya}
\altaffiliation[On leave from ]{Donetsk Insitute for Physics and Technology, National Academy of Sciences of
Ukraine, 83114 Donetsk, Ukraine.}
\affiliation{Institut f\"{u}r Materialwissenschaft, Technische Universit\"{a}t Darmstadt,
D-64287 Darmstadt, Germany }
\date{\today }

\begin{abstract}
We study the penetration of the nonuniform magnetic field, created by a
magnetic dipole with out-of-plane magnetization, into a film heterostructure
composed of a type-II superconductor layer and a soft-magnet layer. In the
framework of the London approach, the energy of the magnetic dipole-vortex
interaction is derived and the critical value of the dipole moment for the
first appearance of a vortex in the superconducting constituent is found for
two cases of the layer ordering, namely when the dipole is located near the
superconducting or, respectively, the magnetic constituent.
\end{abstract}

\pacs{74.78.-w, 74.25.Op}
\maketitle

Recent progress in microfabrication technology has made hybrid structures
composed of superconducting and \textit{ferromagnetic} materials a very
popular object of study. In particular, there were conducted many
experimental and theoretical studies of the structures composed of
superconductors (SC's) and ferromagnetic dots (FD's) (see, for example,
Refs. \cite{rev1,rev2,rev3} and references therein). Due to the large
intrinsic magnetic moment of dots they generate diverse vortex
configurations inside a SC layer~\cite%
{dot1,dot2,dip1,dip2,dotP1,dotP2,dotPM,Erdin05}. Also the dots act as
effective trapping centra for the vortices causing in this way an
enhancement of vortex pinning and increase of the critical magnetic field
and of the critical current of a superconductor~\cite%
{rev2,dip2,dip3,pin2,pin1}.

On the contrary, heterostructures of superconductors and \textit{soft magnets%
} (SM's) are studied for a short time. Soft magnets, such as Permalloy, pure
iron, crioperm, etc., have, as a rule, sufficiently large values of the
relative permeability, very narrow hysteresis loop and possess negligible
remanent magnetization. Nevertheless, they may significantly improve
superconductor performance by effective shielding from the external magnetic
field as well as from the transport current self-field~\cite%
{SM1,SM1a,SM2,SM3,SM3a,SM4}.

\begin{figure*}[tbp]
\includegraphics[width=14cm]{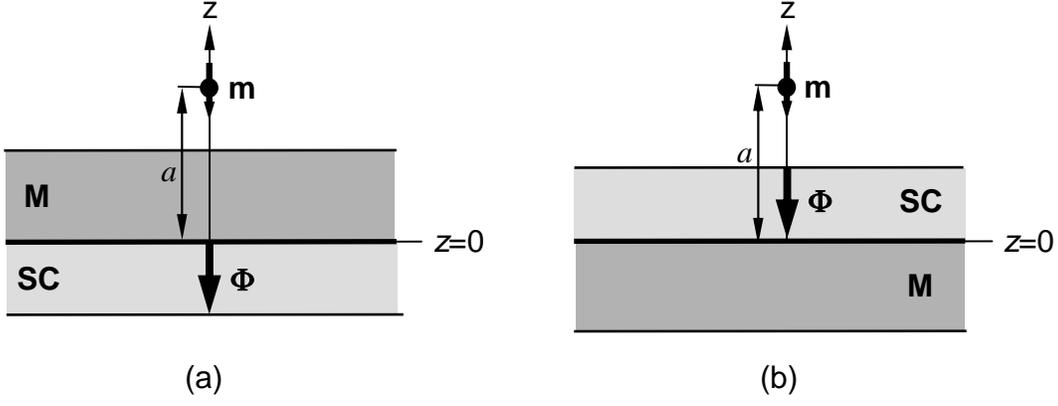}
\caption{Scheme of the first vortex appearance in the bilayer
heterostructure composed of the SC layer and the SM layer exposed by the
field of the point magnetic dipole positioned over the SC layer (a) and over
the SM layer (b).}
\label{f.1}
\end{figure*}

In the present paper we consider a situation which combines both above
structure types. Recently, it was shown theoretically that the magnet sheath
can strongly enhance the Bean-Livingston barrier against the first entry of
nonuniform magnetic field into a type-II superconductor whereas the
condition of the first penetration of the uniform magnetic field is not
influenced by such a sheath~\cite{JPCM,PRBYG}. The similar effect could be
also expected in planar SC-FD heterostructures because of a strong
inhomogeneity of the FD magnetic field. To clear up this question, we study
the penetration of a nonuniform magnetic field, created by a small magnetic
dot, into a \textit{model} film structure of a finite thickness composed of
a type-II SC layer and of a SM layer. Likewise some previous papers~\cite%
{dip1,dip2,dip3}, we represent the dot as a point magnetic dipole (MD)
positioned above the bilayer (Fig.~\ref{f.1}). As was usually the case in
the experiment (see, for example, Ref.~\cite{rev2}), we assume that thin
layers of insulating oxide separate the magnetic and superconducting
constituents of the heterostructure to avoid the proximity effect and
exchange of electrons between them. We consider two cases of the layer
ordering with respect to the MD position, namely when the SM layer is
situated over the SC one (MD-SM/SC configuration, Fig.~\ref{f.1},a) and the
opposite arrangement (MD-SC/SM configuration, Fig.~\ref{f.1},b). Using the
London approach, we study how parameters of the magnet layer~-- the
thickness and the relative magnetic permeability~-- influence the
interaction between the dipole and a single vortex in the SC layer and find
the conditions for the first vortex appearance.

Let us consider a point magnetic dipole (MD) with a moment $\mathbf{m}$,
placed above an infinite film heterostructure consisted of a type-II SC
layer with thickness $d_{\text{S}}$ and a SM layer with thickness $d_{\text{M%
}}$ and a relative magnetic permeability $\mu >1$ (see Fig.~\ref{f.1}). An
interface between the superconductor and the magnet lies in the $z=0$ plane
while the MD is positioned at $(x,y,z)=(0,0,a)$ and is magnetized in the
negative $z$ axis direction. In our consideration we neglect the remanent
magnetization as well as both nonlinear behavior and conductivity of the SM
layer, so that the magnetic induction and the magnetic field in this layer
are connected by the relation $\mathbf{b}_{\text{M}}=\mu \mu _{0}\mathbf{h}_{%
\text{M}}$ and, therefore, a relative permeability is assumed the only
characteristic of a homogeneous, isotropic SM constituent ($\mu _{0}$ is the
permeability of free space). The magnetic induction in the SC layer $\mathbf{%
b}_{\text{S}}$ is described by the London equation~\cite{deGennes} 
\begin{equation}
\mathbf{b}_{\text{S}}+\lambda ^{2}\mathbf{\nabla }\times \left( \mathbf{%
\nabla }\times \mathbf{b}_{\text{S}}\right) =\mathbf{Q,}  \label{e.1}
\end{equation}

\noindent with the London penetration depth $\lambda $ and the source
function $\mathbf{Q}$\ accounting for the presence of vortices in the
superconductor. In the Meissner state of the SC layer the function $\mathbf{Q%
}=0$. When a straight vortex directed perpendicular to the film plane, is
present inside the superconductor, the vector $\mathbf{Q}$\ has only $z$%
-component, which in the geometry of Fig.~\ref{f.1} is 
\begin{equation}
Q_{z}=-L\Phi _{0}\delta \left( \mathbf{\rho }-\mathbf{\rho }_{\text{v}%
}\right) .  \label{e.2}
\end{equation}

\noindent Here $\Phi _{0}$ is the flux quantum, $L$\ is the vorticity which
defines the number of flux quanta trapped by the vortex, and two-dimensional
vector $\mathbf{\rho }_{\text{v}}$ is the vortex position in the film plane.
It is conveniently to represent the magnetic field above the heterostructure
as $\mathbf{H}^{\left( 0\right) }+\mathbf{h}_{+}$, where $\mathbf{H}^{\left(
0\right) }$ is the direct contribution from the dipole and $\mathbf{h}_{+}$
is the field induced above the film by the supercurrent subject to the
influence of the SM layer. The field $\mathbf{H}^{\left( 0\right) }$ has the
customary form 
\begin{equation}
\mathbf{H}^{\left( 0\right) }\left( \mathbf{r}\right) =\frac{1}{4\pi }\frac{3%
\mathbf{r}\left( \mathbf{m\cdot r}\right) -\mathbf{m}\left( \mathbf{r}\cdot 
\mathbf{r}\right) }{\left\vert \mathbf{r}\right\vert ^{5}},  \label{e.3}
\end{equation}

\noindent where $\mathbf{r}\equiv \left\{ x,y,z-a\right\} $ is a position
vector with respect to the dipole. The magnetic field $\mathbf{h}_{+}$, the
magnetic field below the film $\mathbf{h}_{-}$, the magnetic field inside
the magnet layer $\mathbf{h}_{\text{M}}$ and the magnetic induction $\mathbf{%
b}$ in the whole space satisfy the Maxwell equations 
\begin{equation}
\mathbf{\nabla }\times \mathbf{h}=0,\qquad \mathbf{\nabla }\cdot \mathbf{b}%
=0.  \label{e.4}
\end{equation}

We imply the existence of an insulating, nonmagnetic layer of thickness much
less than $\lambda $, $d_{\text{M}}$, $d_{\text{S}}$ and $a$ between the
superconductor and the magnet layer (for example, such a layer was
experimentally observed in MgB$_{2}$/Fe wires~\cite{SM4}). According to this
assumption, the boundary conditions for the normal (n)\ and tangential (t)\
field components, applied on the SC/SM interface, read 
\begin{equation}
b_{\text{S,n}}=\mu _{0}\mu h_{\text{M,n}};\quad b_{\text{S,t}}=\mu _{0}h_{%
\text{M,t}}.  \label{e.5}
\end{equation}

\noindent Also, we apply the conventional boundary conditions~\cite{LL8} on
the outer surfaces of the heterostructure, namely a continuity of the
magnetic induction on the SC/vacuum interface and a continuity of normal
components of the induction as well as tangential components of the magnetic
field on the SM/vacuum interface.

To find the condition for the first vortex appearance, we consider the
excess Gibbs energy of the heterostructure due to presence of the single
vortex in the SC layer, 
\begin{equation}
G=F_{\text{v}}-\dint dV_{\text{MD}}\left( \mathbf{h}_{\text{v}}\cdot \mathbf{%
M}\right) .  \label{e.6}
\end{equation}

\noindent Here $F_{\text{v}}$ is the self-energy of a vortex, the second
term describes the energy of interaction between the vortex and the magnetic
dipole~\cite{LL8}, where the integration is performed over the dipole volume~%
$V_{\text{MD}}$, $\mathbf{h}_{\text{v}}$ is the magnetic field of the vortex
in the dipole position, and $\mathbf{M}=\left\{ 0,0,-\mu _{0}m\right\}
\delta \left( x\right) \delta \left( y\right) \delta \left( z-a\right) $ is
the dipole magnetization. The vortex self-energy takes the form~\cite{PRBYG} 
\begin{equation}
F_{\text{v}}=\frac{1}{2\mu _{0}}\int dV_{\text{S}}\left( \mathbf{b}_{\text{v}%
}\cdot \mathbf{Q}\right) +\frac{1}{2}\int dS_{\text{S}}\left. \left[ \psi _{%
\text{out}}\left( \mathbf{Q}\cdot \mathbf{n}\right) \right] \right\vert _{S_{%
\text{S}}},  \label{e.7}
\end{equation}

\noindent where the integration is performed over the volume $V_{\text{S}}$
and the surface $S_{\text{S}}$ of the SC layer, respectively, $\mathbf{n}$
denotes the outer normal to the surface of the SC region, $\psi _{\text{out}%
} $ is the scalar potential of the vortex magnetic field outside the SC
layer. After minimization of the Gibbs energy~$G$\ with respect to the
vortex position~$\mathbf{\rho }_{\text{v}}$\ the condition $G=0$\ determines
the critical value of the dipole moment $m_{\text{c1}}$\ for the first
vortex appearance in the SC layer.

The free self-energy of the single vortex located in the SC layer is
described in both considered configurations by the expression 
\begin{equation}
F_{\text{v}}=\frac{\left( L\Phi _{0}\right) ^{2}}{4\pi \mu _{0}\lambda }%
\left[ \frac{d_{\text{S}}}{\lambda }\ln \left( \frac{\lambda }{\xi }\right)
+\int_{0}^{\infty }dq\frac{\Delta _{1}}{k^{3}\Delta }\right] ,  \label{e.8}
\end{equation}

\noindent with 
\begin{eqnarray}
\Delta _{1} &=&2\mu \left[ k+q\tanh \left( \frac{kd_{\text{S}}}{2\lambda }%
\right) \right] +\tanh \left( \frac{qd_{\text{M}}}{\lambda }\right) \\
&&\times \left[ \left( \mu ^{2}+1\right) k+2q\tanh \left( \frac{kd_{\text{S}}%
}{2\lambda }\right) \right] ,  \notag
\end{eqnarray}

\begin{eqnarray}
\Delta &=&\mu \left[ k+q\tanh \left( \frac{kd_{\text{S}}}{2\lambda }\right) %
\right] \\
&&\times \left[ k+q\coth \left( \frac{kd_{\text{S}}}{2\lambda }\right) %
\right] +\tanh \left( \frac{qd_{\text{M}}}{\lambda }\right)  \notag \\
&&\times \left[ \left( \mu ^{2}+1\right) kq\coth \left( \frac{kd_{\text{S}}}{%
\lambda }\right) +\mu ^{2}k^{2}+q^{2}\right] .  \notag
\end{eqnarray}

\begin{figure}[tbp]
\includegraphics[width=8.5cm]{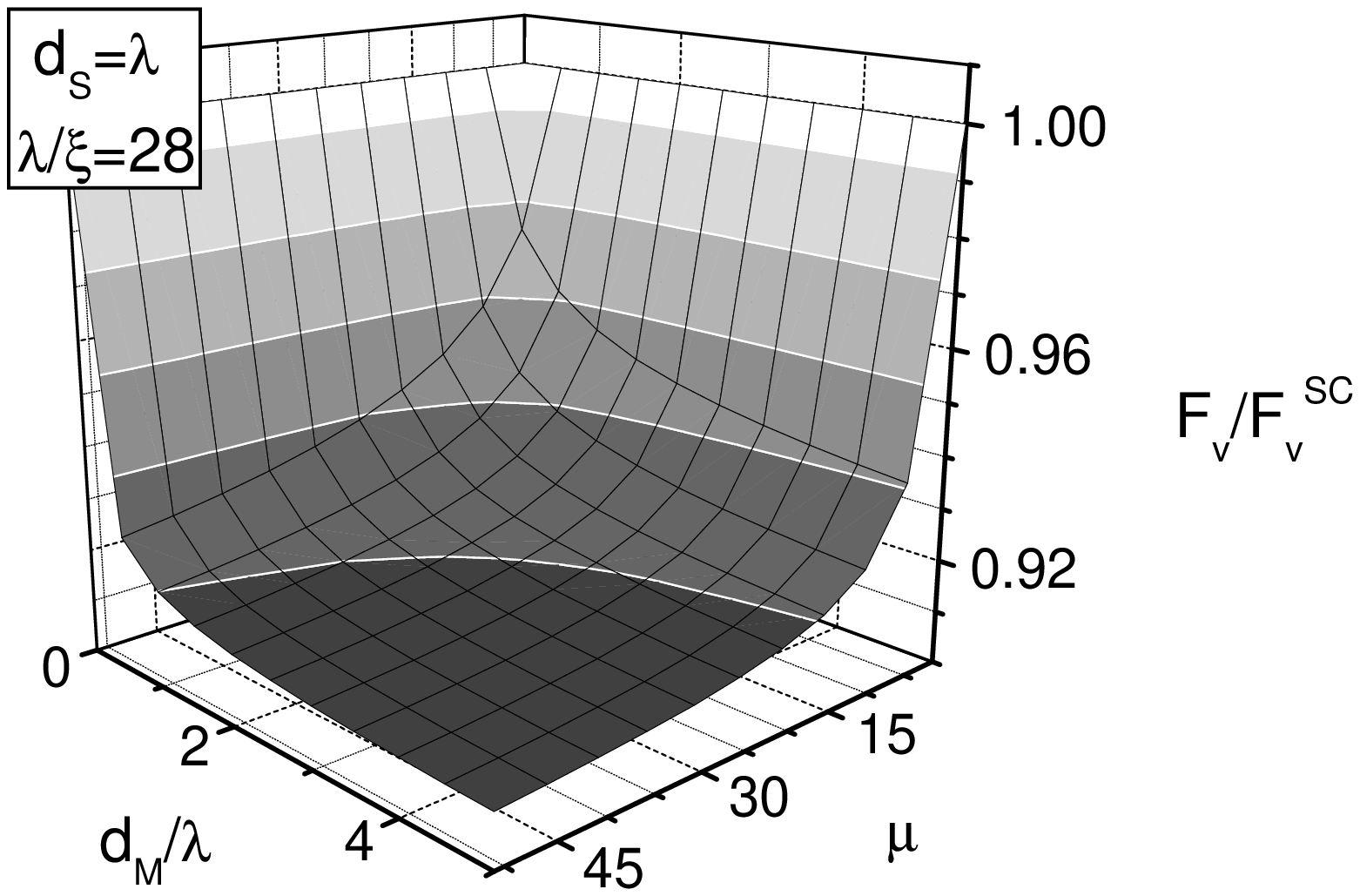}
\caption{The dependence of the vortex free energy $F_{\text{v}}$ on the
thickness $d_{\text{M}}$ and on the relative permeability $\protect\mu $ of
the SM layer.}
\label{f.2}
\end{figure}

\noindent Here $k=\left( 1+q^{2}\right) ^{1/2}$ and $\xi $ is the
superconductor coherence length. The calculated dependence of $F_{\text{v}}$%
\ on $d_{\text{M}}$\ and $\mu $\ normalized on the value of this energy $F_{%
\text{v}}^{\text{SC}}$\ for the unshielded SC film is shown in Fig.~\ref{f.2}
for the case of $d_{\text{S}}=\lambda $\ and $\lambda /\xi =28$. One can see
that due to the presence of SM layer the self-energy of the vortex slightly
decreases. This change reaches the maximum value for the case of $d_{\text{S}%
}\simeq \left( 0.1\div 1\right) \lambda $\ and becomes smaller when $d_{%
\text{S}}\ll \lambda $\ or $d_{\text{S}}\gg \lambda $.

Contrary to the free energy of the vortex, the influence of SM layer on the
interaction between the magnetic dipole and the vortex, situated in the
point with the position vector $\mathbf{\rho }_{\text{v}}$, can be
significant and is different for two considered geometries of the system.
For the MD-SM/SC configuration the energy of the dipole-vortex interaction $%
F_{\text{dv}}=-\dint dV_{\text{MD}}\left( \mathbf{h}_{\text{v}}\cdot \mathbf{%
M}\right) $\ takes the form 
\begin{equation}
F_{\text{dv}}^{\left( \text{I}\right) }=-\frac{L\Phi _{0}m\mu }{2\pi \lambda
^{2}}\int_{0}^{\infty }dq\frac{q\Delta _{2}}{k\Delta }J_{0}\left( \frac{%
q\rho _{\text{v}}}{\lambda }\right) \exp \left[ -\frac{q\left( a-d_{\text{M}%
}\right) }{\lambda }\right] ,
\end{equation}

\noindent where 
\begin{equation}
\Delta _{2}=\frac{k+q\tanh \left( kd_{\text{S}}/2\lambda \right) }{\cosh
\left( qd_{\text{M}}/\lambda \right) },
\end{equation}

\noindent and $J_{0}$\ denotes the Bessel function of the $0$th order. A
crude approximation of this expression has shown that in the limit $\mu \gg
1 $ the energy $F_{\text{dv}}^{\left( \text{I}\right) }$ decreases as $\mu
^{-1}$. For the MD-SC/SM configuration the interaction energy reads 
\begin{equation}
F_{\text{dv}}^{\left( \text{II}\right) }=-\frac{L\Phi _{0}m}{2\pi \lambda
^{2}}\int_{0}^{\infty }dq\frac{q\Delta _{3}}{k\Delta }J_{0}\left( \frac{%
q\rho _{\text{v}}}{\lambda }\right) \exp \left[ -\frac{q\left( a-d_{\text{S}%
}\right) }{\lambda }\right] ,  \label{e.13}
\end{equation}

\noindent with 
\begin{eqnarray}
\Delta _{3} &=&\mu \left[ k+q\tanh \left( \frac{kd_{\text{S}}}{2\lambda }%
\right) \right]  \label{e.14} \\
&&+\tanh \left( \frac{qd_{\text{M}}}{\lambda }\right) \left[ \mu
^{2}k+q\tanh \left( \frac{kd_{\text{S}}}{2\lambda }\right) \right] ,  \notag
\end{eqnarray}%
\noindent and in the limit of $\mu \gg 1$\ has the finite value independent
of $\mu $. Such a behavior of the dipole-vortex interaction energy causes
the different conditions for the first vortex appearance in the SC layer.

Minimizing the Gibbs energy~(\ref{e.6})\ with respect to the vortex position
we obtain that in both geometries the equilibrium position of the vortex is $%
\rho _{\text{v}}=0$, i.e. the vortex is situated exactly under the dipole as
it is shown in Fig.~\ref{f.1}. So, the condition of the first vortex
appearance in the superconductor is determined by the equation 
\begin{equation}
F_{\text{v}}+F_{\text{dv}}\left( \rho _{\text{v}}=0\right) =0.  \label{e.15}
\end{equation}

\begin{figure}[tbp]
\includegraphics[width=8.5cm]{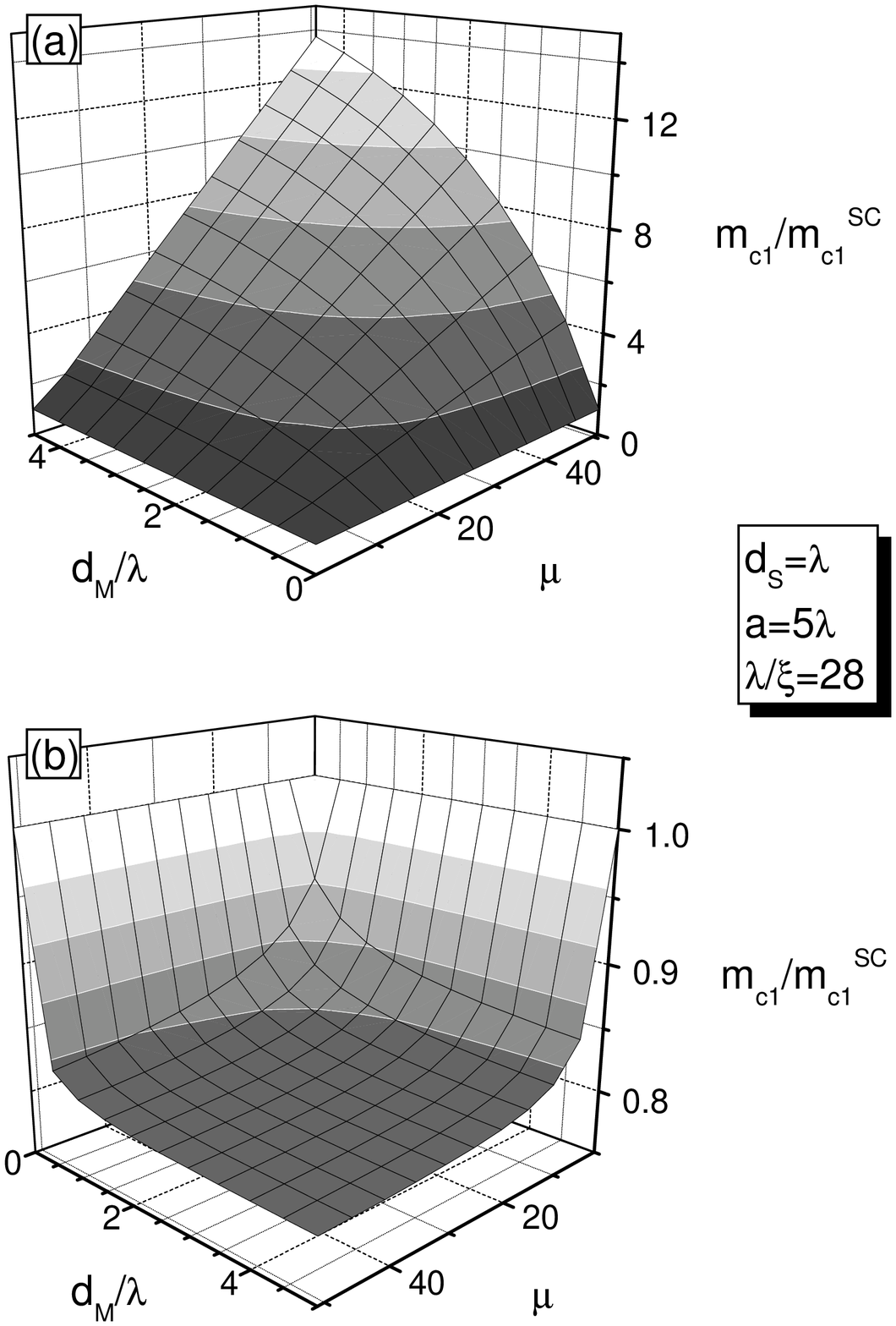}
\caption{The dependence of the dipole magnetic moment $m_{\text{c1}}$ of the
first vortex appearance on the thickness $d_{\text{M}}$ and on the relative
permeability $\protect\mu $ of the SM layer for the configurations shown in
Figs.~\protect\ref{f.1},a and \protect\ref{f.1},b, respectively.}
\label{f.3}
\end{figure}

The calculated dependences of the magnetic moment $m_{\text{c1}}\left( d_{%
\text{M}},\mu \right) $,\ normalized on the value of this moment $m_{\text{c1%
}}^{\text{SC}}$\ for the unshielded SC film, are shown in Fig.~\ref{f.3} for
the heterostructure with $d_{\text{S}}=\lambda $, $a=5\lambda $ and $\lambda
/\xi =28$. One can see that in the case of MD-SM/SC configuration the moment 
$m_{\text{c1}}$ monotonically increases with increase of both $\mu $\ and $%
d_{\text{M}}$ (Fig.~\ref{f.3},a). With increase of $\mu $\ at fixed
thickness $d_{\text{M}}$\ the $m_{\text{c1}}\left( \mu \right) $\ dependence
approximates to the linear one corresponding to the above mentioned limiting
case $\mu \gg 1$. Notice, that such a behavior is qualitatively similar to
the same dependences of the field of first magnetic flux entry into the
magnetically shielded SC filament~\cite{PRBYG} where the local magnetic
field near the place of flux entry is also nonuniform. Therefore, in the
MD-SM/SC configuration the SM layer can significantly prevent the
penetration of vortices into the superconductor preserving the latter in the
Meissner state. The $m_{\text{c1}}\left( d_{\text{M}},\mu \right) $
dependence for the case of MD-SC/SM configuration (Fig.~\ref{f.3},b) is
quite different. The moment $m_{\text{c1}}$\ monotonically decreases with
increase of $\mu $\ and/or $d_{\text{M}}$\ and reaches the finite minimum
value. So, in this case the presence of SM constituent decreases the
magnetic moment $m_{\text{c1}}$\ and promotes to the earlier appearance of
the vortex in the SC layer.

The obtained difference of the dipole-vortex interaction in the considered
cases can be explained in the following way. As it follows from Eq.~(\ref%
{e.6}), the interaction energy $F_{\text{dv}}$\ is determined by the value
of normal component of the magnetic field created by the vortex in the
dipole position. We found that in the MD-SM/SC configuration this field is
significantly reduced ($\mu $\ times in the $\mu \gg 1$\ limit) whereas in
the MD-SC/SM configuration the field of the vortex in the region over the
heterostructure is changed only slightly. Such a difference causes the
corresponding striking change in the interaction.

Notice, that all dependences shown in Figs.~\ref{f.2} and~\ref{f.3} do not
depend explicitly on the value of vorticity~$L$ and, therefore, they will
demonstrate the same behavior with respect to $d_{\text{M}}$\ and $\mu $\
when a vortex (or an antivortex) with arbitrary vorticity is located in the
SC layer.

\begin{figure}[tbp]
\includegraphics[width=8.5cm]{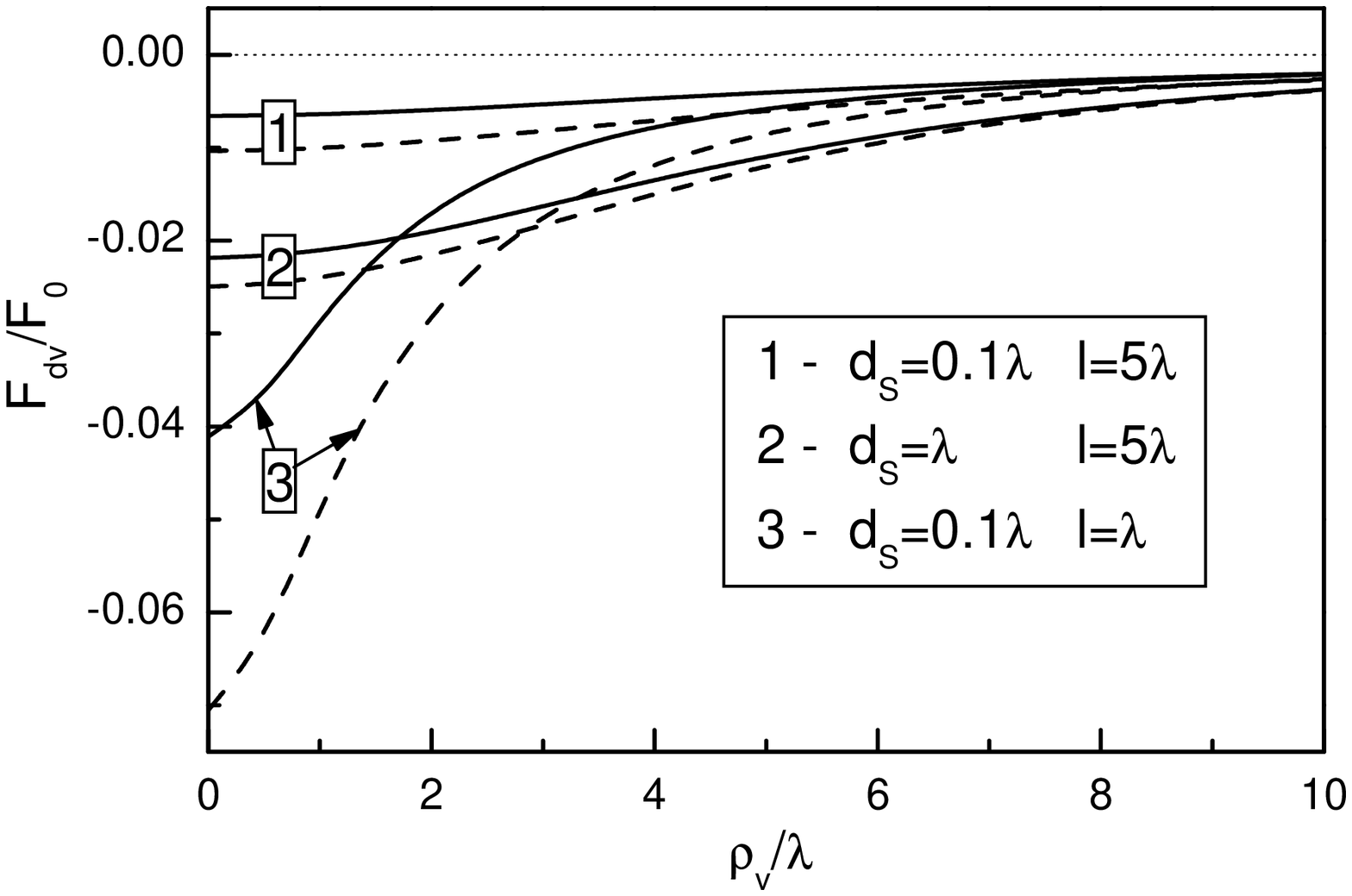}
\caption{The magnetic dipole-vortex interaction energy as a~ function of the
vortex position with respect to the dipole for the heterostructure shown in
Fig.~\protect\ref{f.1},b with the various thicknesses of SC layer and
different distances $l=a-d_{\text{S}}$ between the dipole and the SC layer.
The solid curves indicate the dependences for the structure without the
magnet layer. The corresponding dependences in the presence of the magnet
layer with $d_{\text{M}}=5\protect\lambda $\ and $\protect\mu =50$\ are
shown by dashed curves.}
\label{f.4}
\end{figure}

Though in the geometry of Fig.~\ref{f.3},b the SM layer promotes the earlier
penetration of vortices into the superconductor, another effect occurs in
this case allowing to improve the superconducting properties of the
heterostructure. The dependence of the interaction energy $F_{\text{dv}%
}^{\left( \text{II}\right) }$, measured in units of $F_{0}=L\Phi _{0}m/2\pi
\lambda ^{2}$,\ on the vortex position $\rho _{\text{v}}$\ is shown in Fig.~%
\ref{f.4}\ for the different parameters of the heterostructure. It is
clearly seen that the depth of the potential well for the vortex in the
field of the dipole increases when the SM layer is present. Therefore, in
the case of MD-SC/SM configuration the presence of the SM layer may enhance
the single vortex pinning by the dipole.

In conclusion, we have studied in the framework of the London approach how
the nonuniform magnetic field, created by the point magnetic dipole with a
moment~$\mathbf{m}$, penetrates into the film heterostructure composed of a
type-II SC layer and of a SM layer. We have derived the critical value of
the dipole moment for the first appearance of a vortex and demonstrate that
the presence of SM layer can effectively improve the superconducting
properties of the heterostructure. On the one hand, the SM layer shields the
SC constituent from the dipole field keeping the latter in the Meissner
state when the dipole is situated above the SM constituent. On the other
hand, when the dipole is positioned over the SC constituent the SM layer
enhances the pinning of the vortex by the dipole. Finally, though we have
restricted ourselves to the simplest model when the point dipole creates
only one straight vortex in the superconductor, we suppose that, in more
realistic cases, such as a creation of vortex-antivortex structures in the
superconductor~\cite{rev3,dip2,dotP2} or heterostructures composed of
superconductors and dots of finite sizes~\cite{pin1,dotP1,dotPM,Erdin05},
the similar influence of the soft-magnet layer could also take place.

\begin{acknowledgments}
Stimulating discussions with H.~Rauh and Yu.~Genenko are gratefully
acknowledged. This study was supported by a research grant from the German
Research Foundation (DFG).
\end{acknowledgments}

\end{document}